\documentclass[journal,twocolumn]{IEEEtran}
\usepackage{bm}
\usepackage{xcolor}
\usepackage{mathtools}
\usepackage{epsfig,makeidx,color}
\usepackage{amsmath,amssymb,bbm,enumitem}
\usepackage{cite,graphicx,lipsum}
\usepackage{hyperref}
\hypersetup{
        colorlinks = true,
        citecolor=blue,
}
\usepackage{wrapfig}
\usepackage[demo]{tikzpeople}
\usepackage{draftfigure}
\usepackage{tcolorbox}

\usepackage{booktabs,multirow}
\usepackage{graphicx}
\usepackage{subcaption}
\usepackage[font=small]{caption}

\begin{document}

\title{Interference-Aware Emergent Random Access Protocol for Downlink LEO Satellite Networks}

\author{Chang-Yong Lim,
        Jihong Park, Jinho Choi, Ju-Hyung Lee, Daesub Oh, and Heewook Kim
\thanks{C.-Y. Lim is an independent researcher in Seoul, Korea. J. Park and J. Choi are with the School of IT, Deakin University. VIC 3220, Australia. J.-H. Lee is with the Ming Hsieh Dept. of ECE, USC, Los Angeles, CA 90089, USA. D. Oh and H. Kim are with ETRI, Daejeon 34129, Korea. J. Park is the corresponding author (email:  jihong.park@deakin.edu.au).
This work was supported by the IITP grant funded by MSIT, Korea (No. 2021-0-00719).
}
}

\maketitle

\begin{abstract}
In this article, we propose a multi-agent deep reinforcement learning (MADRL) framework to train a multiple access protocol for downlink low earth orbit (LEO) satellite networks. By improving the existing learned protocol, emergent random access channel (eRACH), our proposed method, coined centralized and compressed emergent signaling for eRACH (Ce2RACH), can mitigate inter-satellite interference by exchanging additional signaling messages jointly learned through the MADRL training process. Simulations demonstrate that Ce2RACH achieves up to 36.65\% higher network throughput compared to eRACH, while the cost of signaling messages increase linearly with the number of users.

\end{abstract}





\vspace{-5pt}
\section{Motivation} \label{Sec:Intro}
A massive constellation of low earth orbit (LEO) satellites is envisaged to be a native component in 6G non-terrestrial networks (NTNs) \cite{Giordani21,Xingqin21}. 
To facilitate this, we focus on random access protocols to coordinate multiple users in downlink LEO satellite networks. Most of current communication protocols assume fixed infrastructure and operational areas, which allows for manual crafting and standardization of access control operations. However, this approach faces challenges in LEO satellite networks due to their non-stationary topology and wide operational areas.
To address these issues, it is promising to use multi-agent deep reinforcement learning (MADRL) to learn protocol operations tailored for specific environments \cite{mota2021emergence,park2023towards}. Despite the non-stationarity, the orbiting movements of LEO satellites create underlying patterns that can be discerned through MADRL. In this regard, the emergent random access channel (eRACH) protocol has recently been proposed \cite{lee2022learning}. In eRACH, MADRL based optimal random access policies are locally performed by each user, without exchanging any information between LEO satellites and users. However, such fully decentralized operations are challenged by inter-satellite interference. From our experiments in Tab.~\ref{tab:perform_comparison}, we observed that interference decreases the average network throughput of eRACH by up to 58.74\%.

\begin{figure}
    \centering
    \includegraphics[width=.45\linewidth]{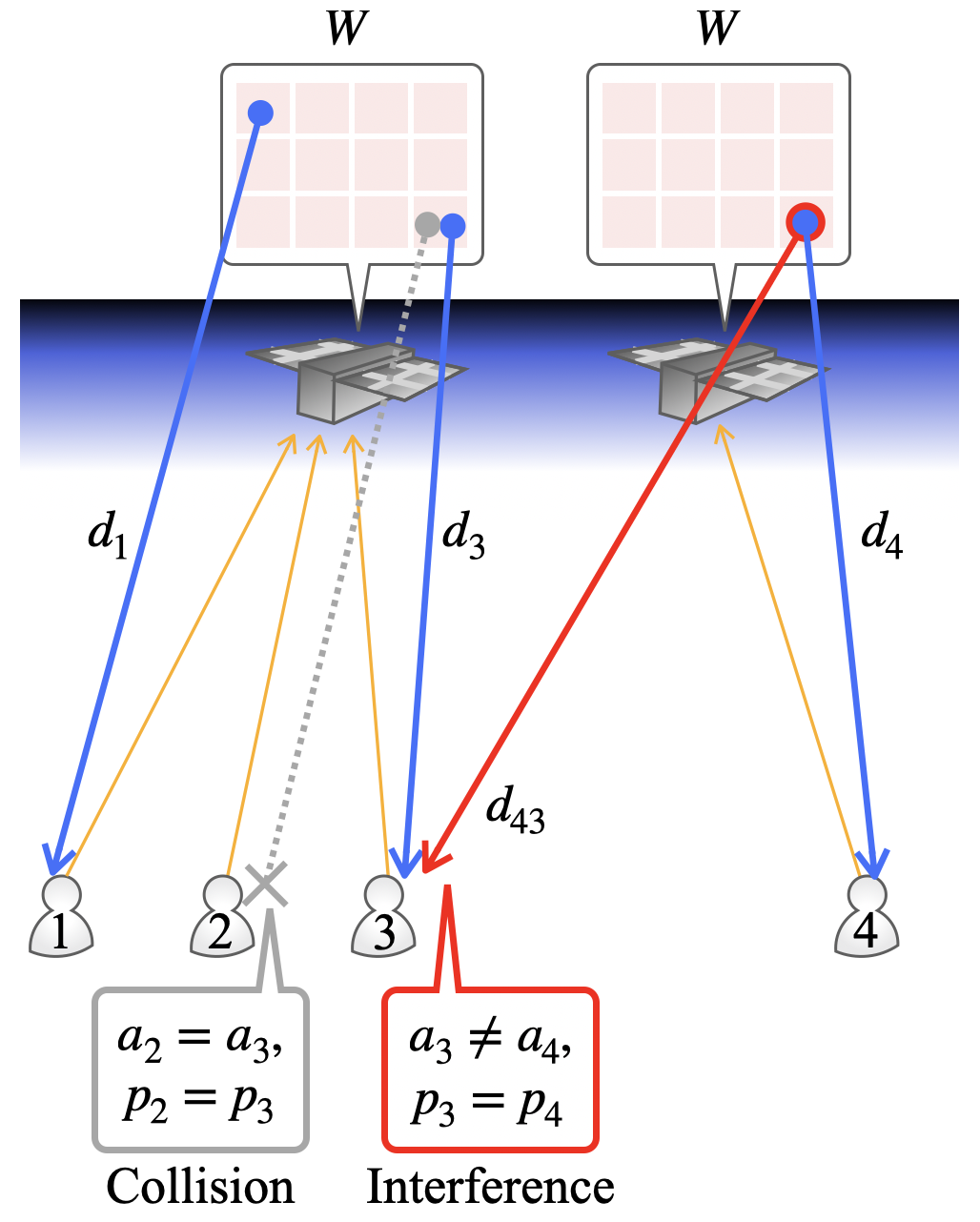}
    \caption{
    Downlink communication after random access with two LEO satellites in the presence of inter-satellite interference.} \vspace{-10pt}
    \label{fig:MADRL_scenario}
\end{figure}

\begin{figure}
    \centering
    \includegraphics[width=.5\textwidth]{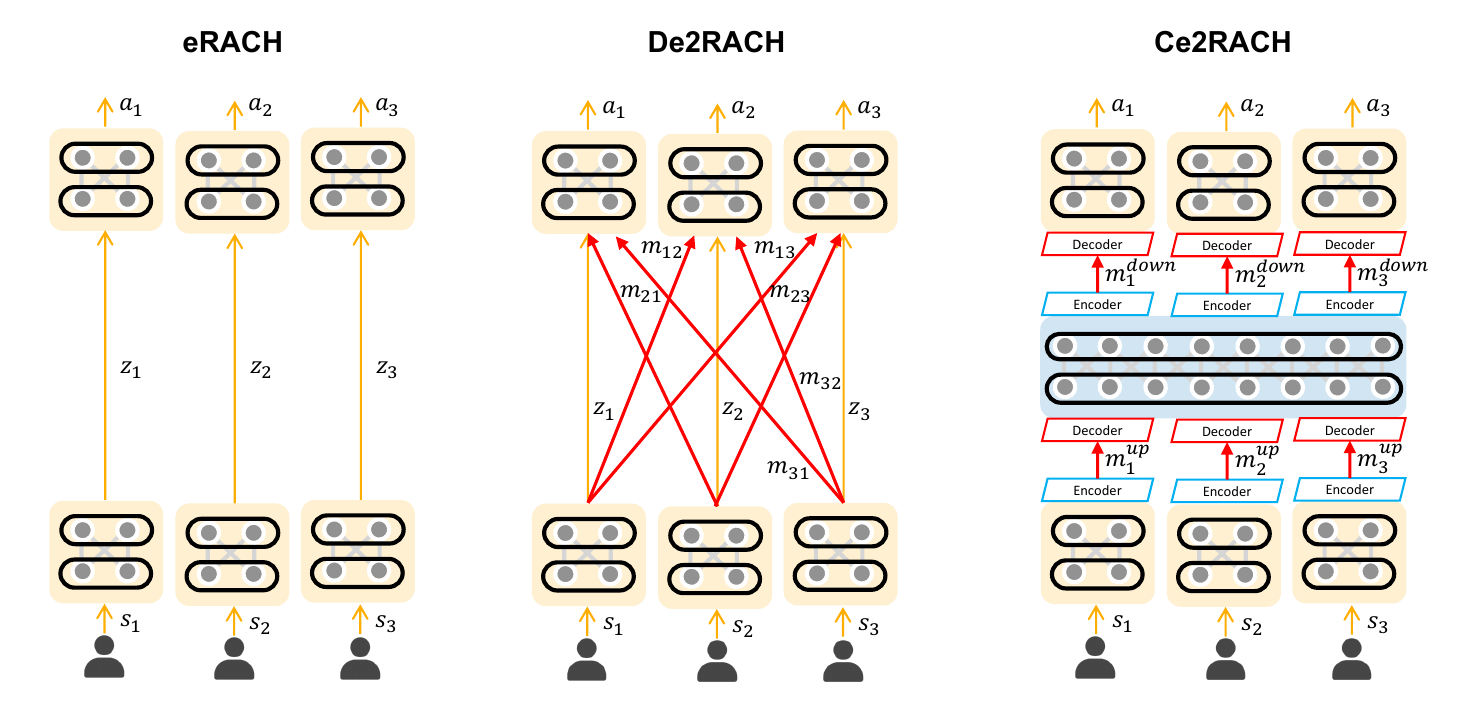}
    \caption{Architectures of eRACH, De2RACH, and Ce2RACH.} \vspace{-5pt}
    \label{Fig:schematic}
\end{figure}
\vspace{-10pt}

\section{Main Result Summary}
To mitigate inter-satellite interference, in this article we propose a novel framework of centralized and compressed emergent signaling for eRACH (Ce2RACH), wherein users exchange additional control signaling messages, inspired by protocol learning frameworks that train signaling messages for specific environments \cite{mota2021emergence}. These learned signaling messages correspond to users’ hidden-layer activations. Existing protocol learning algorithms exchange these activations in a fully decentralized way. This approach, hereafter termed decentralized emergent signaling for eRACH (De2RACH), is certainly not scalable due to high signaling communication overhead. Contrarily, as illustrated in Fig.~\ref{Fig:schematic}, Ce2RACH employs a centralized relay shared by all users, which merges the activations of all users and distributes them to each user, thereby reducing the number of signaling links. Additionally, Ce2RACH compresses signaling messages via deep learning based source coding with an autoencoder, thus reducing the signaling message payload size per link. Simulations demonstrate that Ce2RACH achieves higher network throughput with lower collision probability than eRACH. Furthermore, the additional signaling cost of Ce2RACH linearly increases with the number of users, as opposed to the exponential cost growth of De2RACH.


\begin{table}
    \centering 
    \begin{tabular}{ccc}
    \toprule
        \textbf{Protocol} &  \textbf{Avg. Txpt.} [Mbps] & \textbf{Collision Prob.}\\
        \midrule
       \textcolor{gray}{eRACH w.o. interference} \cite{lee2022learning} & \textcolor{gray}{69.3}  & \textcolor{gray}{5.94}  \\
       \textcolor{gray}{eRACH w. interference} & \textcolor{red}{28.6} & \textcolor{red}{11.9} \\
       De2RACH w. interference & \textbf{38.2} & \textbf{1.98} \\
       Ce2RACH w. interference & \textbf{32.5} & \textbf{9.91} \\
         \bottomrule
    \end{tabular}
    \caption{Network throughput \& collision probability comparisons.}
    \label{tab:perform_comparison}
\end{table}

\begin{figure}
\centering
\subfloat[eRACH. \label{a-Snapshot}]{\includegraphics[width=0.33\linewidth]{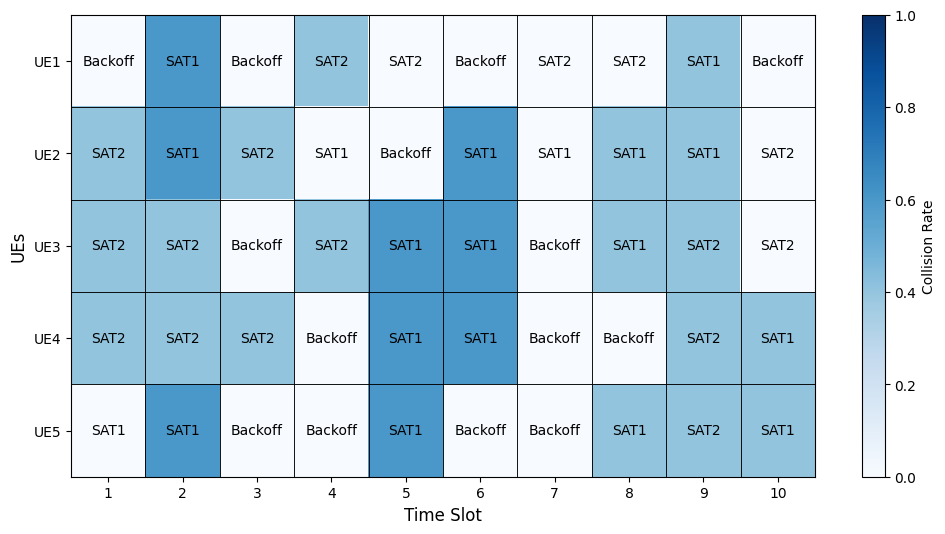}}
\subfloat[De2RACH. \label{b-Snapshot}]{\includegraphics[width=0.33\linewidth]{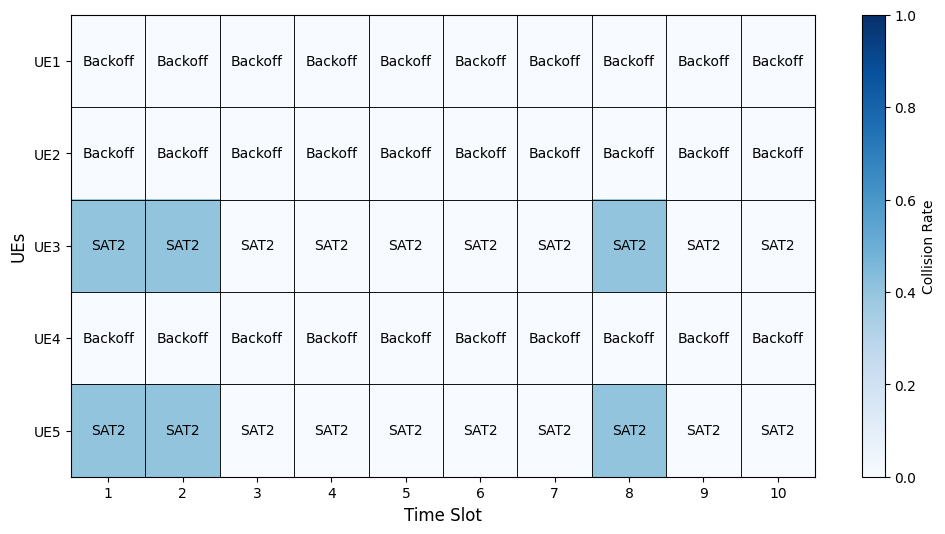}}
\subfloat[Ce2RACH. \label{c-Snapshot}]{\includegraphics[width=0.33\linewidth]{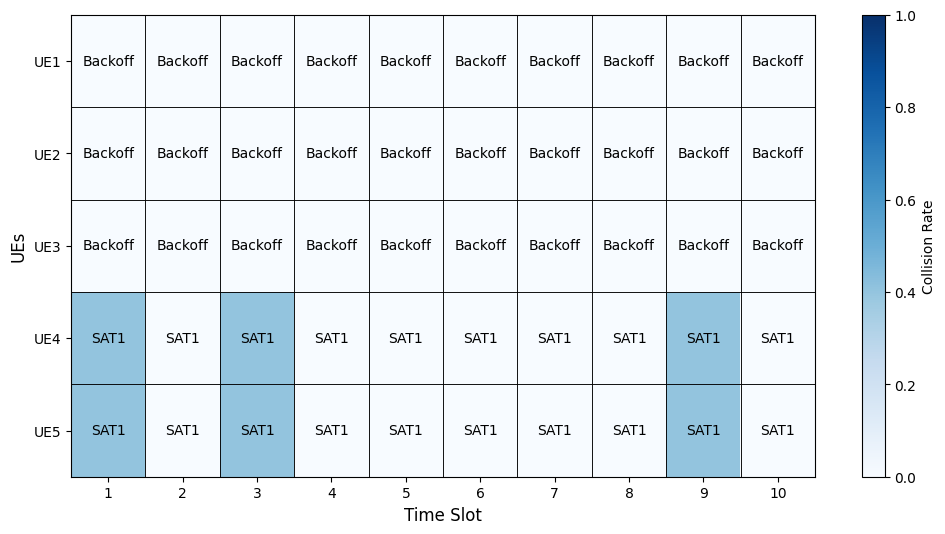}}
\caption{Resource utilization snapshots with respect to collision.}
\label{Fig:utilization}
\end{figure}

\section{System Model}

Consider a downlink network consisting of a set $\mathcal{J}$ of ground users and $K$ LEO satellites orbiting a linear lane with a constant orbiting velocity. At the $n$-th time slot, the $j$-th ground user carries out an action $a_j[n] \in\{0, 1, \cdots, K \}$ for random access,
where $a_j[n] = i$ with $i\neq 0$ implies transmit the user's transmitting a pilot signal to the $i$-th satellite, while $a_j[n] = 0$ represents a backoff operation. Each non-backoff operation is followed by the user's uniformly randomly selecting a pilot signal $p_j[n]\in\{1, \cdots, P \}$ out of $P$ pilot signals.
An event of collision occurs when multiple users transmit the same pilot signal to the same, which is identified using a binary indicator $c_j[n]=1$ if $a_j[n]=a_{j'}[n]$ and $p_j[n]=p_{j'}[n]$ for all $j'\in \mathcal{J}\backslash j$. 

After random access completes without collision, the $j$-th user receives data from the associated satellite with the rate $R_j[n] = \frac{W}{P} \log \left( 1 + \frac{d_j[n]^{-\alpha} }{\sum_{i\in I_j[n]} d_{ij}^{-\alpha} + \sigma^2 / P_t} \right)$, where $d_{ij}^{-\alpha}$ is the distance to the $i$-th interferer. Here, for simplicity, we consider the same transmit power $P_t$ for all users, and assume that interference is aligned with the pilot signal selection,  in the way that the $j$-th user is interfered by the $i$-th satellite when the satellite's associated user uses the same pilot as that of the $j$-th user, as illustrated in Fig. \ref{fig:MADRL_scenario}. Accordingly, a set of interferers $I_j[n]$ seen by the $j$-th user is given as the satellites having their associated users, of which the $k$-th user satisfies: $a_k[n] \neq a_j[n], p_k[n]= p_j[n]$. Given $a_j[n]$, $c_j[n]$, and $R_j[n]$, the user receives the reward $r_j[n]= g(R_j[n] - \rho c_j [n])$, where $\rho>0$ is a constant balancing the importance of rate maximization and collision avoidance, and the function $g(\cdot)$ normalizes the input within the range $[-1,1]$. Other settings follow from \cite{lee2022learning}.

\begin{figure}
    \centering
    \includegraphics[width=.7\columnwidth]{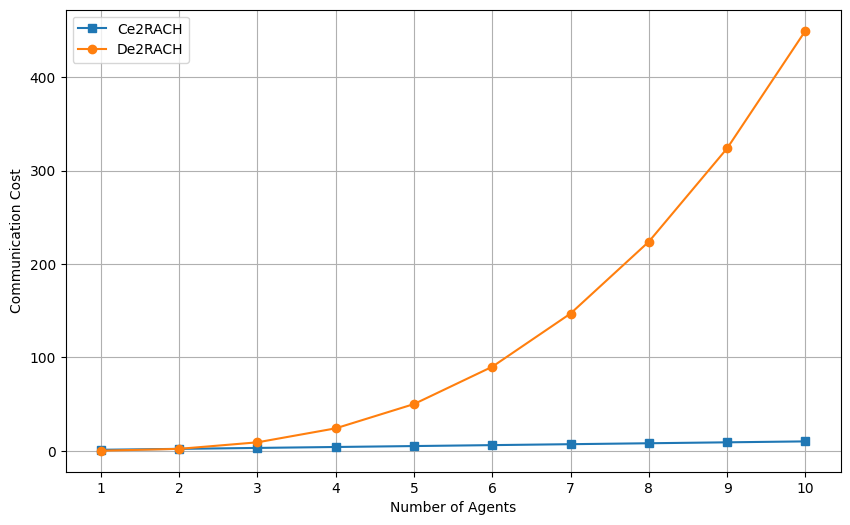}
    \caption{Signaling communication costs of De2RACH and Ce2RACH.}
    \label{Fig:Cost}
\end{figure}

\section{Proposed Methods}
\textbf{eRACH} (baseline) \cite{lee2022learning}: For a given state $s_j[n]$, the $j$-th user's action is represented as $a_j[n] = \theta_j( s_j[n])$, where $\theta_j$ denotes an actor neural network (NN). Dividing $\theta_j$ into upper and lower segments as $\theta_j = [\theta^u_j ; \theta^l_j ]$, it is recast by $a_j[n] = \theta_j^u( z_j[n] )$, where $z_j[n] = \theta_j^l(s_j[n])$. 

\textbf{De2RACH}: Denoting as $\oplus$ the vector concatenation operation, $\theta^l_j$ generates two-fold output $( z_j[n] , \oplus_{k\in\mathcal{J}\backslash j} m_{jk}[n] ) = \theta_j^l(s_j[n])$, where $z_j[n] $ is the input of the upper segment $\theta^u_j$ and $m_{jk}[n]$ is the signaling message from the $j$-th user to the $k$-th user, which follow from \cite{foerster2016learning}. The $j$-th user receives $\oplus_{k\in\mathcal{J}\backslash j} m_{kj}[n] $ from other users, which are fed into $\theta^u_j$, resulting in the action $ a_j[n] = \theta_j^u \left (  z_j[n],   \oplus_{k\in\mathcal{J}\backslash j} m_{kj}[n] \right )$. 

\textbf{Ce2RACH}: 
To reduce the number of signaling links, each user's actor NN is structured as  $\theta_j = [\theta^u_j; \theta_j^{s}; \theta^l_j ]$, and the aggregation $\theta^{s} = \oplus_{j\in\mathcal{J}} \theta_j^{s}$ is shared by all users. This results in $a_j[n] = \theta_j^u( z_j[n], m_j^{\text{dn}}[n])$, where $(z_j[n], m_{j}^{\text{up}}[n]) = \theta_j^l(s_j[n])$ and $m_j^{\text{dn}}[n] = \theta_j^{\text{s}}(\theta_j^l(s_j[n]) )$. To reduce the signaling payload size per link, $m_j^{\text{up}}[n]$ and $m_j^{\text{dn}}[n]$ are encoded using autoencoder NNs as depicted in Fig. \ref{Fig:schematic}. This architecture is inspired from \cite{mota2021emergence} and \cite{nemati2023vq}.






\section{Simulation Results and Conclusion}
We consider a downlink system with 22 LEO satellites and 5 ground users. In the presence of inter-satellite interference, Tab.~\ref{tab:perform_comparison} shows that Ce2RACH improves average network throughput and reduces collision probability by up to 36.65\% and 16.67\%, respectively, compared to eRACH. This is achieved not only by adjusting user-satellite associations but also jointly optimizing backoff operations as visualized in Fig.~\ref{Fig:utilization}. Finally, Fig.~\ref{Fig:Cost} illustrates that the additional signaling cost of Ce2RACH linearly increases with the number of users, as opposed to the exponential cost growth of De2RACH.

Treating De2RACH as a performance upper bound, Ce2RACH achieves sufficiently high network throughput, yet incurs frequent collisions. For future work, it is necessary to balance the throughput-collision trade-off by optimizing $\rho$.

\bibliographystyle{ieeetr}
\bibliography{vtc-spring24.bbl}

\end{document}